\documentclass[iop,revtex4,twocolappendix]{emulateapj}
\usepackage{natbib}
\usepackage[colorlinks=true,urlcolor=blue,citecolor=blue,linkcolor=blue,breaklinks]{hyperref}
\usepackage{amsmath}
\usepackage{etoolbox}

\makeatletter

\patchcmd{\NAT@citex}
  {\@citea\NAT@hyper@{%
     \NAT@nmfmt{\NAT@nm}%
     \hyper@natlinkbreak{\NAT@aysep\NAT@spacechar}{\@citeb\@extra@b@citeb}%
     \NAT@date}}
  {\@citea\NAT@nmfmt{\NAT@nm}%
   \NAT@aysep\NAT@spacechar\NAT@hyper@{\NAT@date}}{}{}

\patchcmd{\NAT@citex}
  {\@citea\NAT@hyper@{%
     \NAT@nmfmt{\NAT@nm}%
     \hyper@natlinkbreak{\NAT@spacechar\NAT@@open\if*#1*\else#1\NAT@spacechar\fi}%
       {\@citeb\@extra@b@citeb}%
     \NAT@date}}
  {\@citea\NAT@nmfmt{\NAT@nm}%
   \NAT@spacechar\NAT@@open\if*#1*\else#1\NAT@spacechar\fi\NAT@hyper@{\NAT@date}}
  {}{}

\begin{document}
\title{ALMA Reveals Transition of Polarization Pattern with Wavelength in HL Tau's Disk}

\author{Ian W. Stephens\altaffilmark{1},  Haifeng Yang\altaffilmark{2}, Zhi-Yun Li\altaffilmark{2}, Leslie W. Looney\altaffilmark{3}, Akimasa Kataoka\altaffilmark{4,5},
Woojin Kwon\altaffilmark{6,7}, Manuel Fern\'andez-L\'opez\altaffilmark{8}, Charles L. H. Hull\altaffilmark{5,9,10},
Meredith Hughes\altaffilmark{11}, Dominique Segura-Cox\altaffilmark{3,12}, Lee Mundy\altaffilmark{13}, Richard Crutcher\altaffilmark{3}, Ramprasad Rao\altaffilmark{14}
}

\altaffiltext{1}{Harvard-Smithsonian Center for Astrophysics, 60 Garden Street, Cambridge, MA, USA;
ian.stephens@cfa.harvard.edu}
\altaffiltext{2}{Astronomy Department, University of Virginia, Charlottesville, VA 22904, USA}
\altaffiltext{3}{Department of Astronomy, University of Illinois, 1002 West Green Street, Urbana, IL 61801, USA}
\altaffiltext{4}{National Astronomical Observatory of Japan, Mitaka, Tokyo 181-8588, Japan}
\altaffiltext{5}{NAOJ Fellow}
\altaffiltext{6}{Korea Astronomy and Space Science Institute, 776 Daedeok-daero, Yuseong-gu, Daejeon 34055, Korea}
\altaffiltext{7}{Korea University of Science and Technology, 217 Gajang-ro, Yuseong-gu, Daejeon 34113, Korea}
\altaffiltext{8}{Instituto Argentino de Radioastronom\'ia, (CCT-La Plata, CONICET; CICPBA), C.C. No. 5, 1894,Villa Elisa, Argentina}
\altaffiltext{9}{National Astronomical Observatory of Japan, NAOJ Chile Observatory, Alonso de C\'ordova 3788, Office 61B, 7630422, Vitacura, Santiago, Chile}
\altaffiltext{10}{Joint ALMA Observatory, Alonso de C\'ordova 3107, Vitacura, Santiago, Chile}
\altaffiltext{11}{Van Vleck Observatory, Astronomy Department, Wesleyan University, Middletown, Connecticut 06459, USA}
\altaffiltext{12}{Max-Planck-Institute for Extraterrestrial Physics, Giessenbachstrasse 1, D--85748 Garching, Germany}
\altaffiltext{13}{Department of Astronomy, University of Maryland, College Park, MD 20742, USA}
\altaffiltext{14}{Institute of Astronomy and Astrophysics, Academia Sinica, 645 N. Aohoku Place, Hilo, HI 96720, USA}

\begin{abstract}
The mechanism for producing polarized emission from protostellar disks at (sub)millimeter wavelengths is currently uncertain. Classically, polarization is expected from non-spherical grains aligned with the magnetic field. Recently, two alternatives have been suggested. One polarization mechanism is caused by self-scattering from dust grains of sizes comparable with the wavelength, while the other mechanism is due to grains aligned with their short axes along the direction of radiation anisotropy. The latter has recently been shown as a likely mechanism for causing the dust polarization detected in HL~Tau at 3.1\,mm. In this paper, we present ALMA polarization observations of HL~Tau for two more wavelengths: 870\,$\mu$m and 1.3\,mm. The morphology at 870\,$\mu$m matches the expectation for self-scattering, while that at 1.3\,mm shows a mix between self-scattering and grains aligned with the radiation anisotropy. The observations cast doubt on the ability of (sub)millimeter continuum polarization to probe disk magnetic fields for at least HL Tau. By showing two distinct polarization morphologies at 870\,$\mu$m and 3.1\,mm and a transition between the two at 1.3\,mm, this paper provides definitive evidence that the dominant (sub)millimeter polarization mechanism transitions with wavelength. In addition, if the polarization at 870\,$\mu$m is due to scattering, the lack of polarization asymmetry along the minor axis of the inclined disk implies that the large grains responsible for the scattering have already settled into a geometrically thin layer, and the presence of asymmetry along the major axis indicates that the HL Tau disk is not completely axisymmetric. 
\end{abstract}

\subjectheadings{protoplanetary disks -- polarization -- stars: formation -- stars: protostars  -- dust  }

\maketitle



\section{Introduction}
Spinning dust grains in the interstellar medium are expected to align with their short axes perpendicular to the magnetic field as a result of radiative torques \citep{Lazarian2007,Andersson2015}. Given this alignment, the polarization of thermal dust emission at millimeter and submillimeter wavelengths is expected to be perpendicular to the magnetic field. Therefore, polarimetric observations at these wavelengths have frequently been used to determine the magnetic field morphology in the plane of the sky in the interstellar medium, especially in star-forming regions. The morphology of magnetic fields has been studied via dust polarization over a wide range of scales: from Galactic scales \citep[kpc; e.g.,][]{Stephens2011} all the way down to protostellar envelope scales \citep[100s -- 1000\,s\, au; e.g.,][]{Girart2006,Rao2009,Stephens2013,Hull2014}. Magnetic fields are thought to play a crucial role in the accretion process for protostellar disks via magnetorotational instability \citep[e.g.,][]{Balbus1998} and/or disk-winds \citep[e.g.,][]{Blandford1982}. Nevertheless, the magnetic field morphology remains observationally poorly constrained in the circumstellar environment.

The earliest attempts to detect polarization in circumstellar disks found polarization typically perpendicular to the major axis of the disk \citep{Tamura1995,Tamura1999}. However, these observations were at $\sim$2000\,au resolution, so they did not resolve the disks; it is certainly possible that the polarized emission could come from protostellar envelopes. \citet{Hughes2009,Hughes2013} made the first attempts to resolve submillimeter polarization within a disk, but such efforts resulted in non-detections, putting upper limits of the linear polarization fraction, $P=\sqrt{Q^2+U^2}/I$, at $\sim$1\% when averaged over the telescope beam (where $I$, $Q$, and $U$ are Stokes parameters). Most recently, polarization has been detected toward a handful of disks and candidates disks: the Class~0 disk candidate IRAS~16293--2422 \citep{Rao2014}, the disk of Class~I/II source HL~Tau \citep{Stephens2014b,Kataoka2017}, the Class~0 disk of L1527 \citep{SeguraCox2015}, the Class~0 disk candidate of NGC~1333 IRAS~4A \citep{Cox2015}, the Herbig~AE late-stage protoplanetary disk HD~142527 \citep{Kataoka2016b}, and the disk candidate of the high-mass protostar Cepheus~A~HW2 \citep{FL2016}. Polarization toward disks have also been detected at mid-infrared wavelengths of 8.7, 10.3, and 12.5\,$\mu$m \citep{LiDan2016,LiDan2017}. However, polarized emission at mid-infrared wavelengths can occur due to absorption, emission, and sometimes scattering, causing difficulty in interpreting the polarization morphology.
\begin{figure}[ht!]
\begin{center}
\includegraphics[width=0.95\columnwidth]{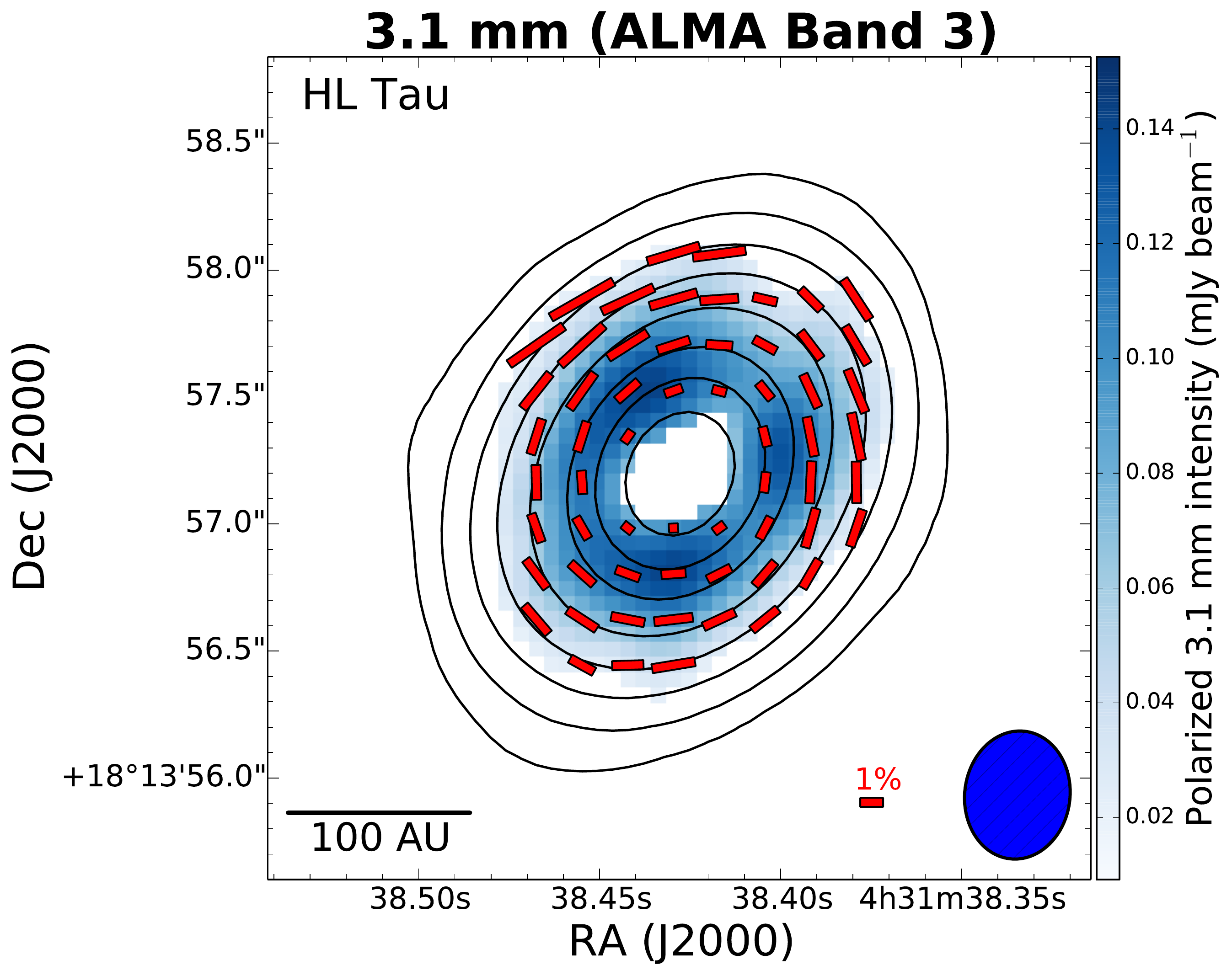}\\
\includegraphics[width=0.95\columnwidth]{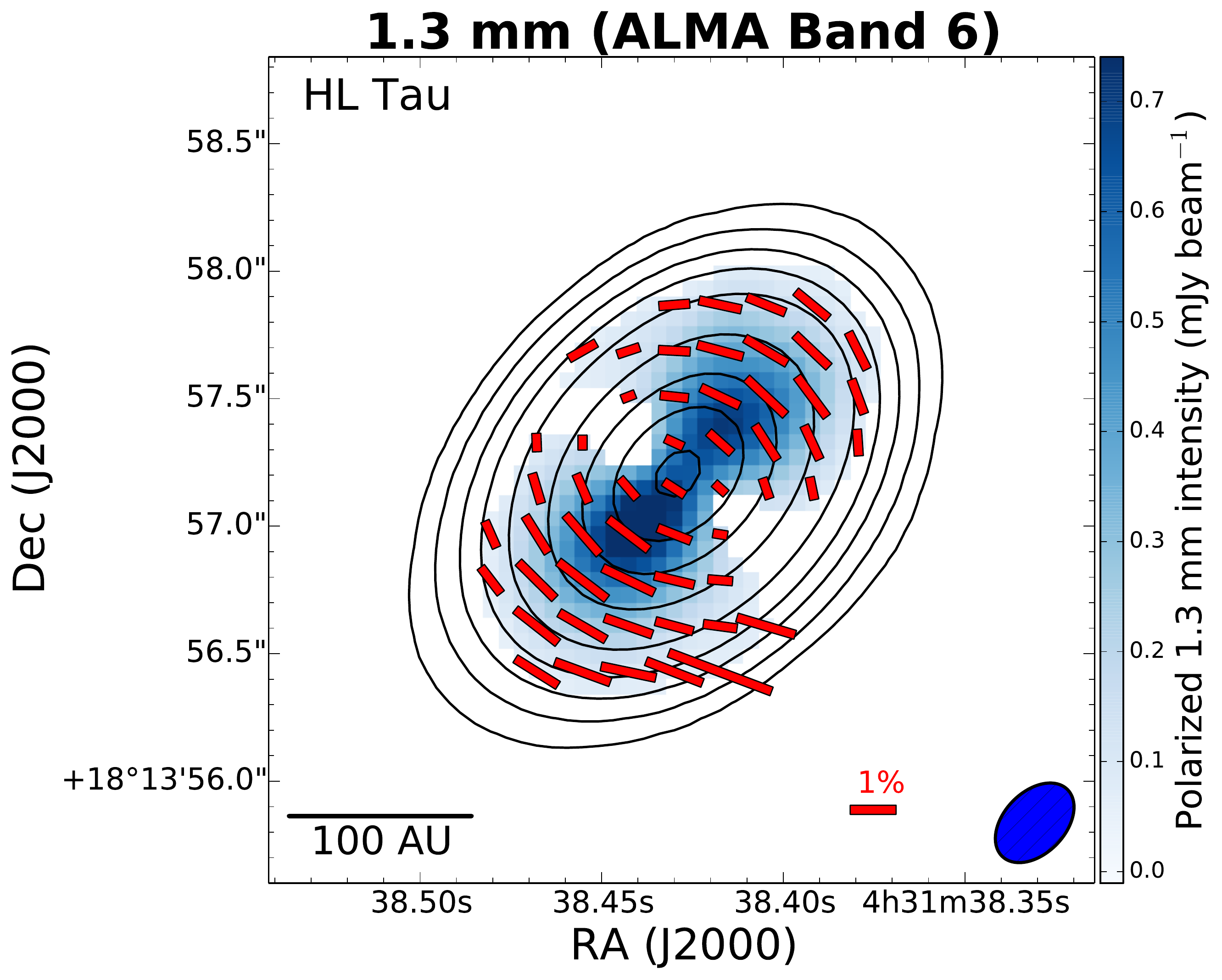}\\
\includegraphics[width=0.95\columnwidth]{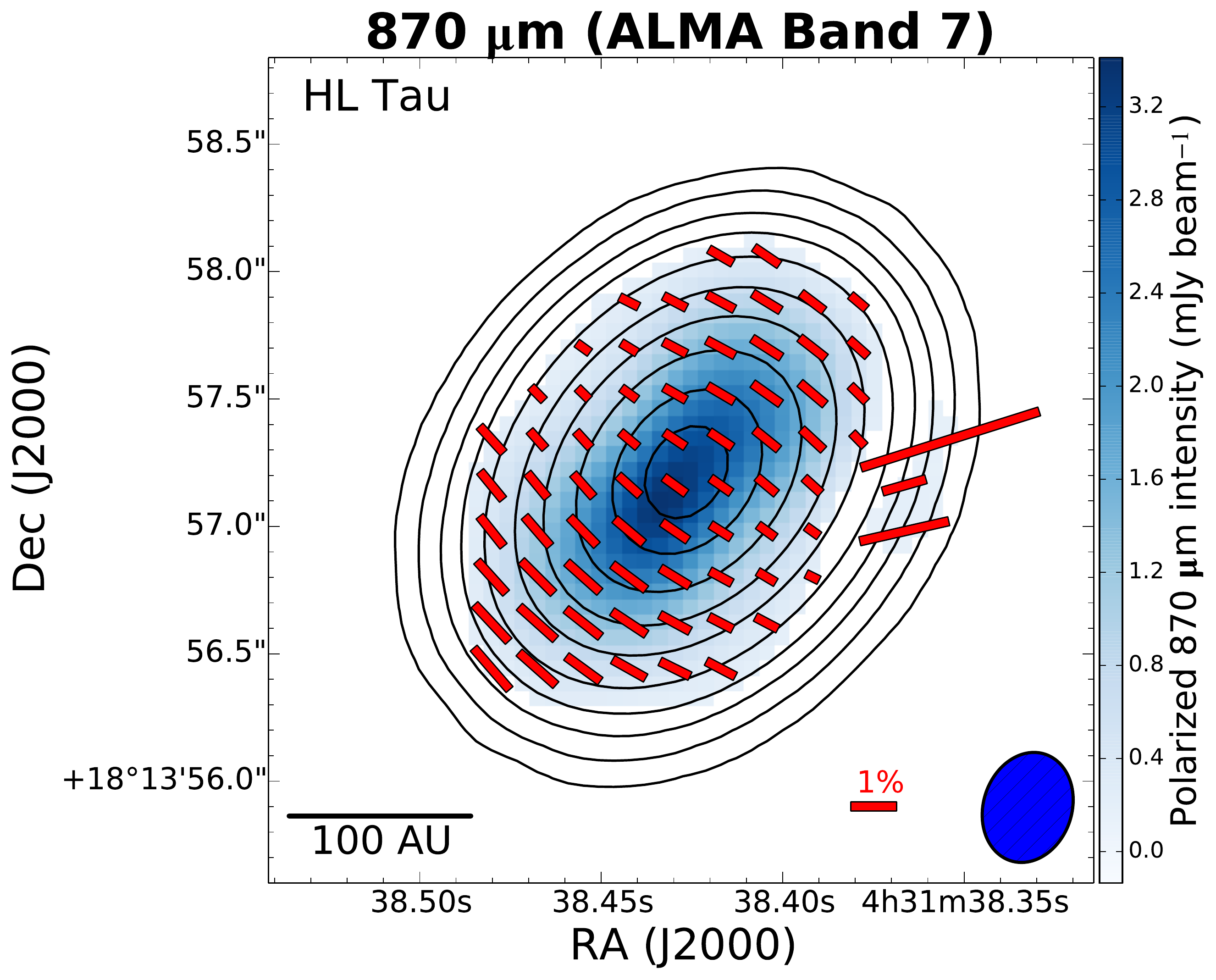}
\end{center}
\caption{ALMA polarimetric observations at 3.1\,mm (top, \citealt{Kataoka2017}), 1.3\,mm (middle), and 870\,$\mu$m (bottom), where the red vectors show the $>$3$\sigma$ polarization morphology (i.e., vectors have not been rotated). Vector lengths are linearly proportional to $P$. The color scale shows the polarized intensity, which is masked to only show 3$\sigma$ detections. Stokes $I$ contours in each panel are shown for [3,\,10,\,25,\,50,\,100,\,200,\,325,\,500,\,750,\,1000]\,$\times$\,$\sigma_I$, where $\sigma_I$ is 44, 154, and 460\,$\mu$Jy\,bm$^{-1}$ for 3.1\,mm, 1.3\,mm, and 870\,$\mu$m, respectively.
}
\label{pol_images} 
\end{figure}


Despite these detections, the polarization morphologies usually were not consistent with what would be expected from magnetically aligned dust grains. In particular, \citet{Stephens2014b} used the Combined Array for Research in Millimeter-wave Astronomy (CARMA) to measure the 1.25~mm polarization morphology in HL~Tau. The morphology was inconsistent with grains aligned with the commonly expected toroidal magnetic fields (polarization/E-field vectors distributed radially in the disk). Instead, the E-vectors were oriented more or less along the minor axis of the disk. \citet{Kataoka2015,Kataoka2016a} and \citet{Yang2016a} suggested that the polarization morphology is actually consistent with that expected from self-scattering \citep[also see][]{Pohl2016,Yang2017}. Indeed, several disks where polarization is detected show consistency with the polarization morphology expected from self-scattering rather than grains aligned with the magnetic field. However, except for the ALMA observations of HD~142527 \citep{Kataoka2016b} and HL~Tau \citep{Kataoka2017}, the published observations are too coarse to resolve more than a few independent beams across the disk, making it difficult to distinguish between scattering and other polarization mechanisms. 

The high-resolution ALMA observations of HD~142527 by \citet{Kataoka2016b} resolved polarization for many 10s of independent resolution elements across the disk. The polarization was radial throughout most of the disk, which is expected for grains aligned with a toroidal field, but toward the edges the morphology changed from radial to azimuthal, which is more consistent with scattering. Models in \citet{Kataoka2016b} found that scattering can broadly reproduce the features observed in parts of the disk -- especially where the polarization orientations change sharply -- but not everywhere. A complete understanding of this interesting case is still missing. 


HL~Tau is one of the brightest Class~I/II disks in the sky at (sub)millimeter wavelengths, and thus the polarization morphology can be determined at high resolution with reasonable integration times. \citet{Kataoka2017} followed up on the \citet{Stephens2014b} observations with 3.1\,mm observations of HL~Tau. Surprisingly, they found that the polarization morphology was azimuthal, which suggests grains aligned with their long axes perpendicular to the radiation field, as predicted by \citet[][also see \citealt{LazarianHoang2007}]{Tazaki2017}. Henceforth, we will call this grain alignment mechanism ``alignment with the radiation anisotropy."

The very different polarization morphologies observed at 1.25\,mm with CARMA \citep[0$\farcs$6 resolution,][]{Stephens2014b} and 3.1\,mm with ALMA \citep[0$\farcs$4 resolution,][]{Kataoka2017} suggest that the morphology of the polarization emission is strongly dependent on the wavelength. The CARMA observations poorly constrained the 1.25\,mm polarization morphology since they only resolved polarization for $\sim$3--4 independent beams across HL Tau. This paper presents ALMA observations at both 1.3~mm and 870\,$\mu$m at resolutions of 0$\farcs$3 and 0$\farcs$4, respectively.

\section{Observations}
With the ALMA Cycle 4 program 2016.1.00162.S (PI: I. Stephens), we observed polarized emission toward HL~Tau at 870\,$\mu$m (Band~7) and at 1.3\,mm (Band~6) on 2016 December 4
and 2017 July 12, respectively. The ALMA configurations for 870~$\mu$m and 1.3\,mm observations were C40-4 and C40-5, respectively. At 870~$\mu$m, the correlator was tuned to four 1.75~GHz spectral windows at center sky frequencies of 336.5, 338.6, 348.5, and 350.5\,GHz, and for 1.3~mm at center frequencies of 224, 226, 240, and 242\,GHz. For both wavelengths, the bandpass and flux calibrator was J0510+1800, the phase calibrator was J0431+1731, and the polarization calibrator was J0522--3627. 


We used the delivered calibrated uv-data and performed imaging with CASA version 4.7.2. We used the CASA \texttt{tclean} task using a Briggs weighting parameter of robust=1. After a quick clean, we performed a series of phase-only self calibration, and images were cleaned after each self calibration iteration with progressively shorter solution intervals. We first performed several rounds of self calibration to the Stokes $I$ image only, and then did self calibration a final round of self calibration on all four Stokes parameters. Images were primary-beam corrected. We also include 3.1\,mm (Band~3) data which were from \citet{Kataoka2017}, except we re-imaged in the identical way described above. The synthesized beams for 3.1\,mm, 1.3\,mm, and 870\,$\mu$m were 0$\farcs$51$\times$0$\farcs$41 with a position angle (measured counterclockwise from north) of --67$\fdg$6; 0$\farcs$37$\times$0$\farcs$24 at --44$\fdg$1; and 0$\farcs$44$\times$0$\farcs$35 at --17$\fdg$9, respectively. Given a distance to HL~Tau of $\sim$140~pc \citep{Rebull2004}, the spatial resolution is 64, 42, and 55~au, respectively.


The ALMA instrumental error on $P$ is expected to be about 0.1\%. In cases where the measured $P$ rms was less than this value, we used 0.1\% as our error. The polarization intensity and $P$ was de-biased via the method described in \citet{Hull2014} and \citet{Hull2015}.


\section{Results}


Figure~\ref{pol_images} shows the observed polarization morphology for HL~Tau for ALMA at 3.1\,mm, 1.3\,mm, and 870\,$\mu$m. The polarization morphology at 3.1\,mm is azimuthal, with a void in the center about the size of the beam. This void likely occurs due to beam-averaging an azimuthal polarization morphology within the beam. The 3.1\,mm polarization morphology is broadly consistent with the pattern expected for emission from grains aligned with radiation anisotropy.\footnote{We will postpone a detailed discussion of the caveats of this interpretation to a future publication.} For 870\,$\mu$m, the polarization morphology is extremely uniform and parallel with the minor axis of the disk. This morphology is predicted for Rayleigh scattering for a geometrically thin (dust) disk \citep{Yang2017}. Toward the west edge of the disk, there is polarization detected that is not aligned with the rest of these vectors. This emission is about the size of the beam and could possibly be a spurious 3$\sigma$ detection.

The 1.3\,mm morphology has slightly more structure but appears to be a superposition of those at 3.1\,mm and 870\,$\mu$m. Notably, along the disk's minor axis, the 3.1\,mm vectors are perpendicular to the axis while the 870\,$\mu$m vectors are parallel. At 1.3\,mm, these cancel, and thus polarization is significantly undetected along the minor axis. On the contrary, along the disk's major axis, 3.1\,mm and 870\,$\mu$m vectors are both perpendicular, and $P$ is highest on average for this line along the disk.

\begin{figure*}[ht!]
\begin{center}
\includegraphics[width=2\columnwidth]{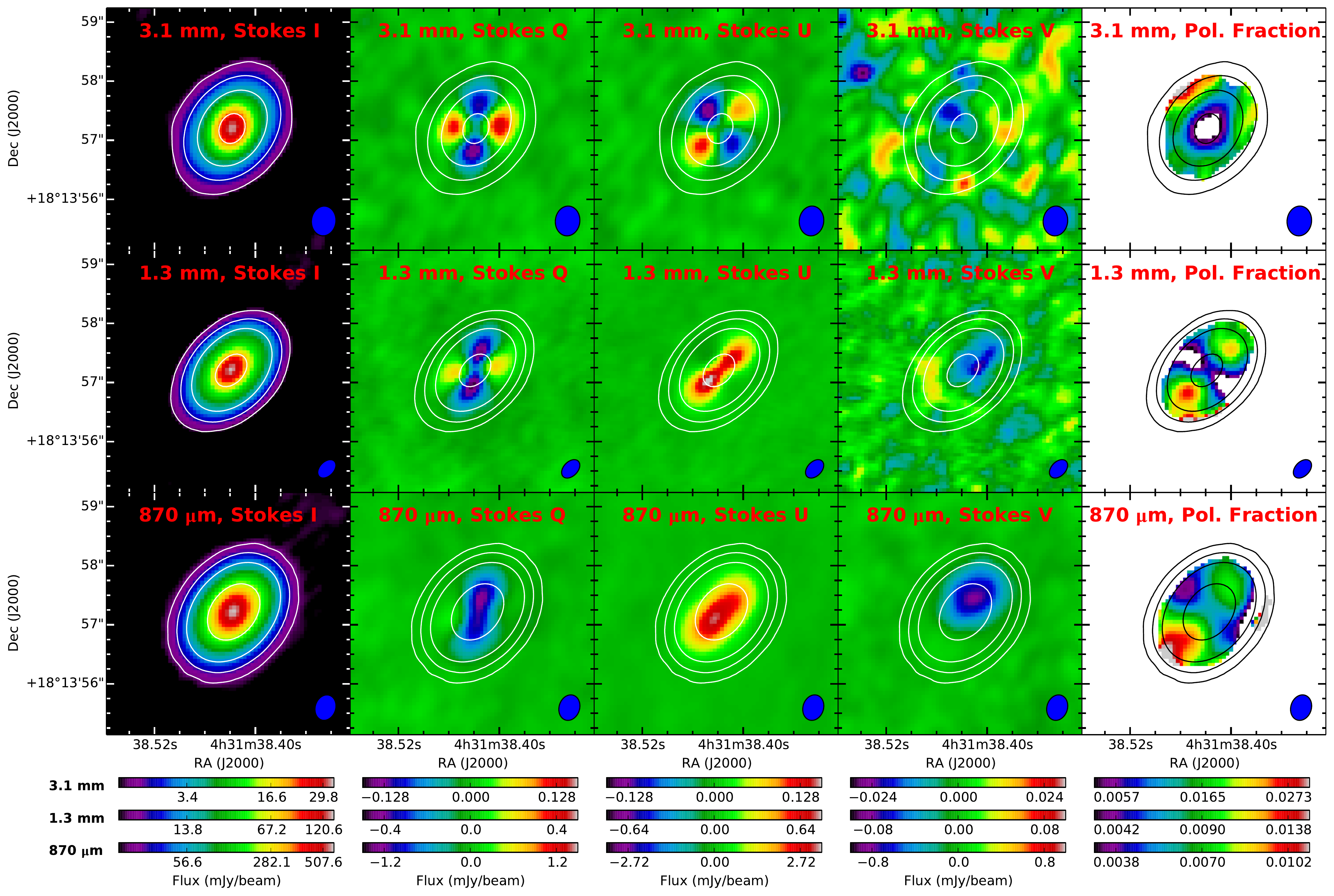}
\end{center}
\caption{ALMA observations of the Stokes parameters $IQUV$ and the linear polarization fraction $P$ for 3.1\,mm (top), 1.3\,mm (middle), and 870\,$\mu$m (bottom). Stokes $I$ contours for a particular wavelength is shown in each panel for [4,\,25,\,100,\,500]$\times\sigma_I$, where $\sigma_I$ is given in Figure~\ref{pol_images}.
}
\label{stokes_images} 
\end{figure*}

The transition of polarization morphologies from one wavelength to another is also apparent in the Stokes~$Q$ and $U$ images, as seen in Figure~\ref{stokes_images}. Specifically, at 3.1\,mm the Stokes~$Q$ image shows negative regions from north to south and positive regions from east to west. At 1.3\,mm, the negative regions are much brighter than the positive regions, and at 870\,$\mu$m, the positive regions almost completely disappear. Stokes~$U$ shows two positive and two negative regions at 3.1\,mm. The negative blobs disappear at 1.3\,mm, while the positive regions connect, and at 870\,$\mu$m, the positive regions turn into a single, large region.

For the Stokes $V$ images (Figure~\ref{stokes_images}), 3.1\,mm appears to be purely noise. For both 1.3\,mm and 870\,$\mu$m, Stokes~$V$ has a negative flux blob offset to the top-right of the Stokes $I$ intensity, and this blob is significantly brighter in 870\,$\mu$m. Although circular polarization can in principle be produced by scattering of linearly polarized light off non-spherical grains \citep[e.g.,][]{Tazaki2017}, Stokes~$V$ has not been well characterized for ALMA, so its detection could be spurious and will not be discussed in detail in this paper. 

The central polarization vectors for 1.3\,mm and 870\,$\mu$m are 55$\fdg$4\,$\pm$0$\fdg$96 and 55$\fdg$3\,$\pm$0$\fdg$41, respectively. The major axis of HL Tau's disk is 138$\fdg$02\,$\pm$\,0$\fdg$07 \citep{ALMA2015}, indicating the polarization vectors are $\sim$7$^\circ$ from being perpendicular to the major axis. We would expect the polarization to be perfectly perpendicular to the major axis. The reason for this small 7$^\circ$ discrepancy is uncertain, but could indicate asymmetry in the disk.

The right panels of Figure~\ref{stokes_images} show $P$ for each wavelength. For 1.3\,mm and 870\,$\mu$m, there is a slight asymmetry for $P$ along the major axis of the disk, with the southeast more polarized than the northwest. For 1.3\,mm, the absolute difference between the two $P$ peaks along the major axis is 0.0013\,$\pm$\,0.0014, which is not statistically significant. The asymmetry for 870\,$\mu$m is more apparent, although the southeast part of the disk does not peak (i.e., it is still increasing at the edge of the detected polarized emission). The $P$ peak toward the northwest is 0.0077\,$\pm$\,0.0010. Drawing a line from the northwest peak through the center of the disk, at the same radius on the southeast side, $P$~=~0.0094\,$\pm$\,0.0010. $P$ rises to $\sim$0.0115 at the edge of the disk. Therefore, the $P$ asymmetry along the major axis for 870\,$\mu$m is $\sim$0.002--0.004. These asymmetries are relatively moderate, but the fact that they appear both at 1.3\,mm and 870\,$\mu$m suggests they could be a real feature. If we consider total polarization $P_T = \sqrt{Q^2+U^2+V^2}/I$, the asymmetry is similar; for 1.3\,mm, it is 0.0012\,$\pm$\,0.0014, while for 870$\mu$m it is $\sim$0.001--0.004.

\citet{Yang2017} showed that asymmetry along the major axis is not expected for scattering-induced polarization in an axisymmetric disk, so axisymmetry must be broken if the polarization is really due to scattering even though the total intensity (Stokes $I$) is highly axisymmetric. On the other hand, the polarization along the minor axis {\it is} expected to be asymmetric if the dust disk is optically thick and has a significant angular width (for a cartoon illustration, see Figure 6 in \citealt{Yang2017}). HL Tau disk is known to be optically thick at 870\,$\mu$m \citep{Carrasco2016}, but no asymmetry is detected along the minor axis. The lack of asymmetry for $P$ along the optically thick minor axis implies that that the large grains responsible for scattering at 870\,$\mu$m are already settled into a thin layer \citep{Yang2017}. Evidence for dust settling in HL Tau was also found in \citet{Kwon2011,Kwon2015a} based on modeling of the millimeter dust continuum, and in \citet{Pinte2016} based on the shape of gaps in dust continuum images





\begin{figure}[ht!]
\begin{center}
\includegraphics[width=1\columnwidth]{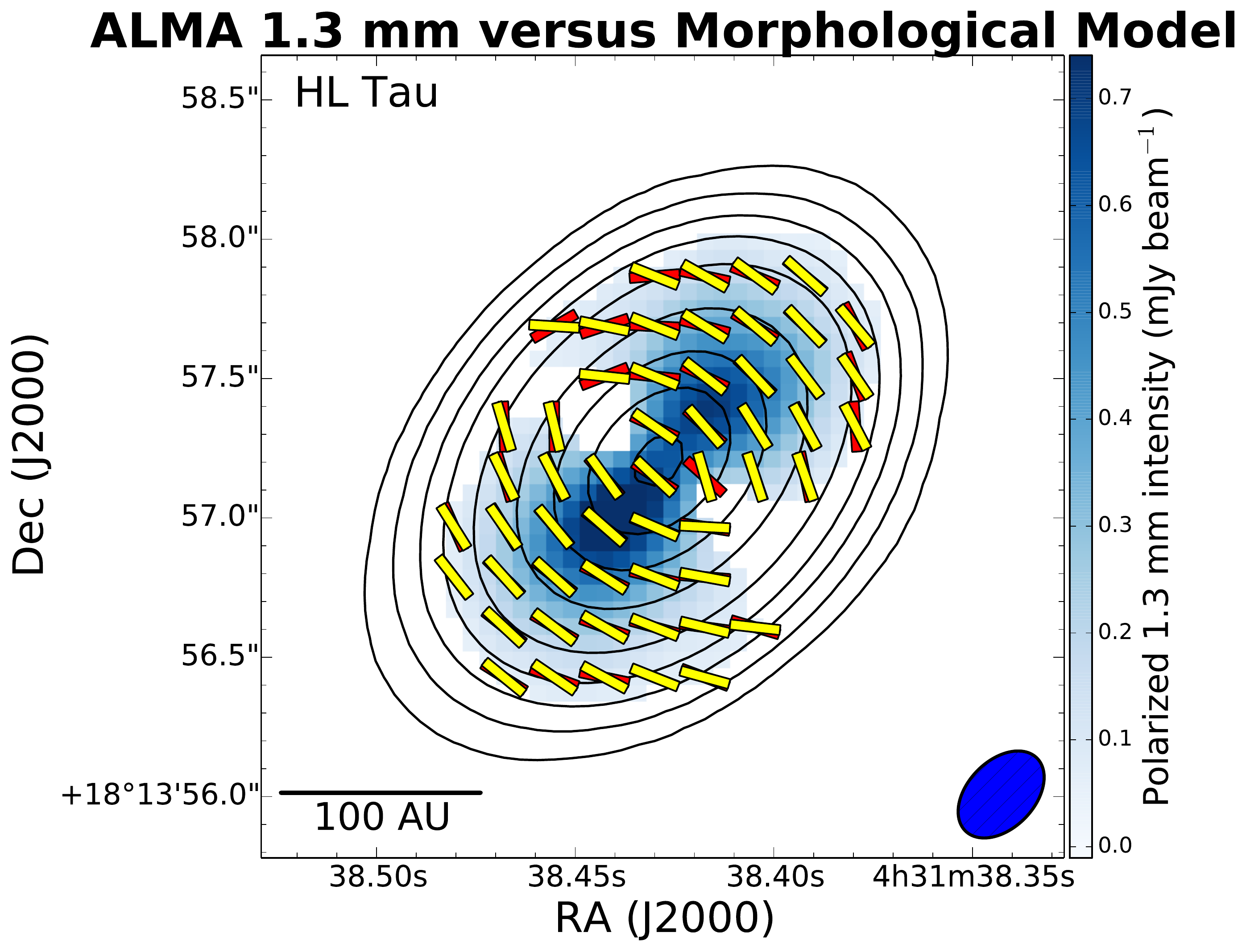}
\end{center}
\caption{Morphological 1.3\,mm model for a 50--50 mix of azimuthal and uniform patterns is shown with yellow vectors, which are overlaid on top of the observed polarization morphology (red vectors). The contours are Stokes~$I$, with the levels given in Figure~\ref{pol_images}.
}
\label{model_image} 
\end{figure}

\section{1.3\,mm Morphological Model of HL~Tau}
For illustrative purposes, we have also developed two simple models for the 1.3\,mm data: one for the polarization orientation and the other for the spatial distribution of the polarized intensity.

The polarization orientation model is a simple 50--50 mix of a purely azimuthal pattern (mimicking the pattern observed at 3.1\,mm) and a uni-directional pattern along the minor axis (an idealization of the pattern observed at 870\,$\mu$m). We created the model by averaging the Stokes $Q$ and $U$ of the two patterns separately. Figure~\ref{model_image} shows the model compared with the observations. The two match remarkably well, which strengthens the notion that the 1.3\,mm emission is a transition case between 3.1\,mm and 870\,$\mu$m.

\begin{figure}[ht!]
\begin{center}
\includegraphics[width=1\columnwidth]{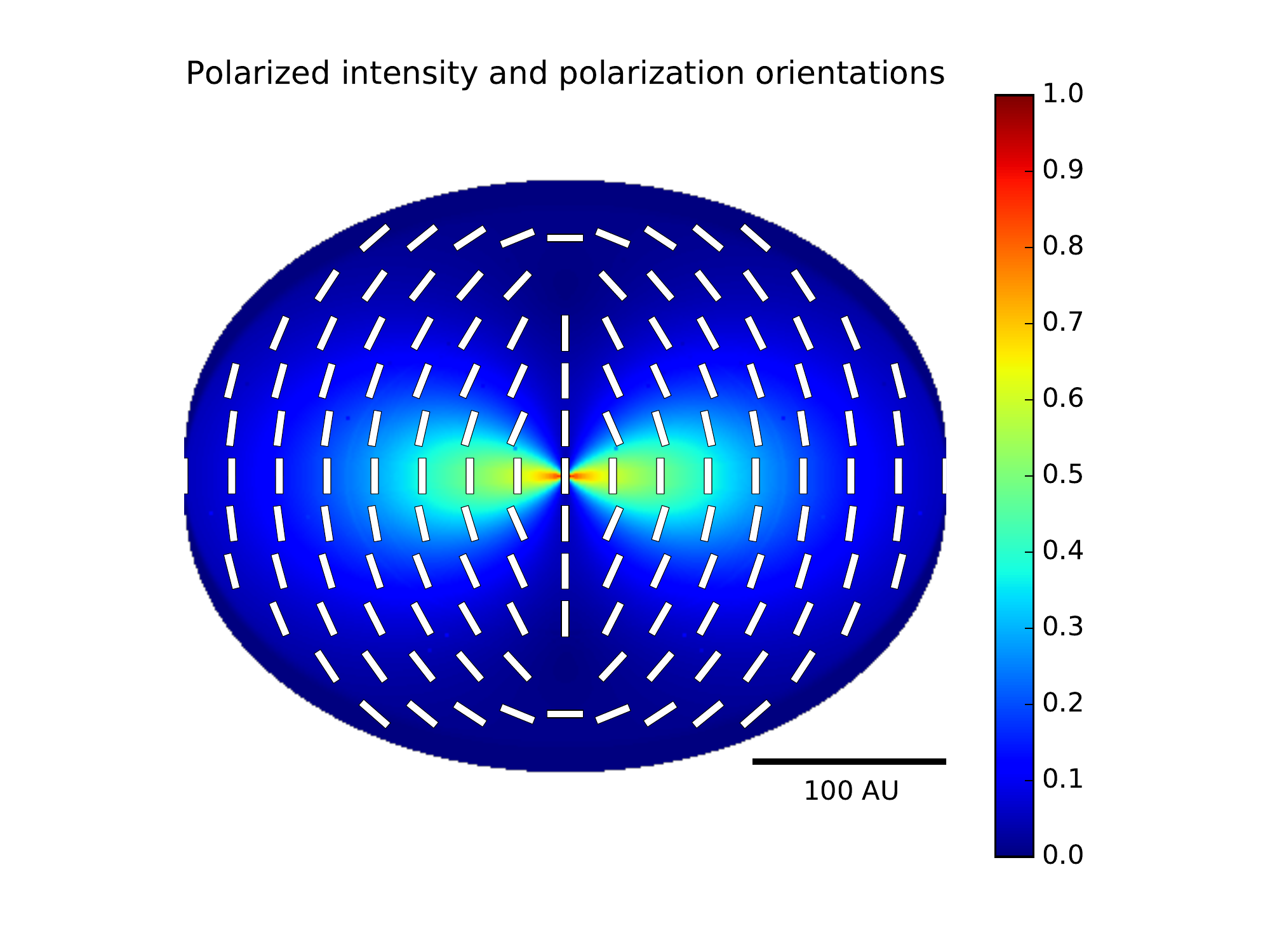}
\end{center}
\caption{Distribution of polarized intensity with polarization vectors superposed for an illustrative disk model where the polarization is produced by a combination of direct emission and scattering by azimuthally aligned grains. Color scale has been normalized to have a peak of 1.
}
\label{calculation_image} 
\end{figure}

The polarized intensity model is based on a disk model similar to Model~A of \citet{Yang2017}, but with a reduced scale height for the dust layer. The grains are assumed to be oblate spheroids with their shortest axes aligned radially, as expected in the case of alignment with the radiation anisotropy. The intrinsic polarization fraction of the non-spherical grains is set to 1\%, and the effective grain size is 30\,$\mu$m. The direct emission and scattering by such grains are computed under the electrostatic approximation (see \citealt{Yang2016b} for details). The resulting polarization pattern is shown in Figure~\ref{calculation_image} and broadly matches the observed pattern, especially the dumbbell-shaped distribution of the polarized intensity and the morphology of the polarization vectors. This model is not unique and does not match all details of the data; a detailed modeling will have many uncertainties in terms of dust properties and disk structure and is beyond the scope of this paper. Nevertheless, it does provide an illustration that the observed pattern at 1.3\,mm can plausibly be explained with a combination of direct emission and scattering by aligned grains. 
%

\section{Summary}
\citet{Kataoka2017} suggested that the polarization emission at 3.1\,mm for HL~Tau is caused by grain alignment via radiation anisotropy. In this paper, we present polarization images at two additional wavelengths. At 870\,$\mu$m the polarization morphology is uniform, which is consistent with polarization due to self-scattering. At 1.3\,mm, the polarization morphology appears to be a mix of the polarization morphologies at 3.1\,mm and 870\,$\mu$m, rather than purely self-scattering, as suggested in other studies \citep[e.g.,][]{Kataoka2015,Kataoka2017,Yang2016a}. These observations are the first to show a clear transition in a disk's (sub)millimeter polarization morphology with wavelength. The observations also provide a clear confirmation of two different mechanisms causing polarization emission at (sub)millimeter wavelengths. These observations are not consistent with polarized emission from grains aligned with a toroidal magnetic field. 

We are able to reproduce the morphology with a morphological model and a theoretical calculation. We also find tentative evidence that the linear polarization fraction $P$ is asymmetric along the major axis of the disk, which would suggest that HL~Tau is not perfectly axisymmetric. Furthermore, the lack of asymmetry along the minor axis at the optically thick wavelength of 870$\mu$m may indicate that the large grains responsible for the scattering-induced polarization have already settled into a geometrically thin layer, which provides an independent check on the dust settling inferred from the shape of the gaps in HL Tau disk.

By observing high-resolution polarization of HL Tau at three different wavelengths, we have the ability to decipher which polarization mechanisms are occurring at different wavelengths. Without observing polarization of HL~Tau at 3.1\,mm and 870\,$\mu$m, the polarization morphology at 1.3\,mm would have been difficult to interpret. Understanding and precisely modeling polarization in a protostellar disk probably require multi-wavelength, high-resolution polarimetric data.

\acknowledgements
H.F.Y. is supported by an ALMA SOS award. Z.Y.L. is supported in part by NASA NNX14AB38G and NSF AST-1313083 and 1716259. A.K. acknowledges support by JSPS KAKENHI grant number JP15K17606. W.K. was supported by Basic Science Research Program through the National Research Foundation of Korea (NRF-2016R1C1B2013642).
 This paper makes use of the following ALMA data: ADS/JAO.ALMA\#2016.1.00115.S and ADS/JAO.ALMA\#2016.1.00162.S. ALMA is a partnership of ESO (representing its member states), NSF (USA) and NINS (Japan), together with NRC (Canada); MOST and ASIAA (Taiwan); and KASI (Republic of Korea), in cooperation with the Republic of Chile. The Joint ALMA Observatory is operated by ESO, AUI/NRAO and NAOJ. The National Radio Astronomy Observatory is a facility of the National Science Foundation operated under cooperative agreement by Associated Universities, Inc.




\end{document}